\documentclass[preprint,12pt]{elsarticle}

\usepackage{amssymb}
\usepackage{amsmath}
\usepackage{amsthm}
\usepackage{ latexsym}
\newcommand{{\bfkappa}}{\mbox{\boldmath${\kappa}$\unboldmath}}
\newcommand{{\bfg}}{\mbox{\boldmath$g$\unboldmath}}
\newcommand{{\bfv}}{\mbox{\boldmath$v$\unboldmath}}
\newcommand{{\bfk}}{\mbox{\boldmath$k$\unboldmath}}
\newcommand{{\bff}}{\mbox{\boldmath$f$\unboldmath}}
\newcommand{{\bfF}}{\mbox{\boldmath$F$\unboldmath}}
\newcommand{{\bfA}}{\mbox{\boldmath$A$\unboldmath}}

\journal{}

\begin{document}

\begin{frontmatter}

\title{Isotope Frequency Shifts of BH Local\\ Pair Modes in Crystalline Silicon}

\author[label1]{Lucio Andrade}
\address[label1]{Departamento de F\'isica, Facultad de Ciencias, Universidad Nacional Aut\'onoma de M\'exico, M\'exico, 04510, D. F., M\'exico  }
\ead{lab@ciencias.unam.mx}
\begin{abstract}
Isotope frequency shifts of local pair modes due to B$^{10}$ (B$^{11}$) substitutional and H (D) interstitial in crystalline silicon are calculated as a many body problem with a simple model and found to be in agreement with experiment. A comparison of the obtained results is made with other theoretical calculations.
\end{abstract}


\end{frontmatter}

\section{Introduction}
Since almost three decades ago considerable interest has been shown in the vibrational spectroscopy of the BH local pair modes, in crystalline silicon. However, it appears that at least one aspect of this spectroscopy, namely, isotope frequency shifts, is still not sufficiently understood [1-6].

Experimental information, at room temperature, includes Raman investigations of the BH complex in silicon [5,7,8]. Two boron local pair modes, due to B substitutional and H interstitial, with frequencies at 680 cm$^{-1}$ B$^{10}$ and at 652 cm$^{-1}$ B$^{11}$ have been found. It is interesting to point out that, boron local pair mode frequencies are of the same order of magnitude as boron local single mode frequencies in crystalline silicon  643 cm$^{-1}$ B$^{10}$ and  620 cm$^{-1}$ B$^{11}$  [9]. These local single mode frequencies had also been measured by infrared in 644 cm$^{-1}$ B$^{10}$ and  620 cm$^{-1}$ B$^{11}$ [10,11]. Note also that, boron local pair mode isotope shift 28 cm$^{-1}$ is similar to boron local single mode isotope shift 24 cm$^{-1}$. Moreover, from these experiments it was concluded that both boron local pair mode frequencies are independent of whether B is joined with H or D [5,7,8].

Previous infrared [12,13] and Raman [8] investigations, at room temperature, had also revealed a line at 1875 cm$^{-1}$ which had been ascribed to hydrogen local pair mode of the BH couple in crystalline silicon. The corresponding line for deuterium had as well been observed at 1360 cm$^{-1}$ [13,14]. Hence, hydrogen local pair mode frequencies also are of the same order of magnitude as hydrogen local single mode frequencies in crystalline silicon 1998 cm$^{-1}$ H [15] and 1449 cm$^{-1}$ D [16]. Here, in addition, note that hydrogen local pair mode isotope shift now is 515cm$^{-1}$and this value is similar to hydrogen local single mode isotope shift which is 549 cm$^{-1}$. Further fine absorption measurements, at low temperature, for the hydrogen (deuterium) local pair mode have also been made [17]. In this last delicate infrared study, the H (D) local pair mode frequency was found to lie at 1905.2 cm$^{-1}$ (1393.9 cm$^{-1}$) for the B$^{10}$H (B$^{10}$D) complex and at 1904.4 cm$^{-1}$ (1390.6 cm$^{-1}$) for theB$^{11}$H ($B^{11}$D) pair. Both these lines $B^{10}$H and B$^{11}$H for the BH complex correspond to that report by Stavola et. al. at 1903 cm$^{-1}$ [18]. The splittings (small isotope frequency shifts) of the hydrogen and deuterium lines, with respect to the two boron isotopes, are respectively 0.8 cm$^{-1}$ and 3.3 cm$^{-1}$. Therefore, it was clear from these experiments, that hydrogen (deuterium) local pair mode frequency is dependent of whether H(D) is joined with B$^{10}$ or B$^{11}$ [12,13,17,18]. It should be noticed here that this last inference is contrary to that above mentioned concerning the interaction between boron and hydrogen atoms in crystalline silicon.

If chemical bonds between impurities and between these impurities with the lattice exist then it is impossible to attribute local modes to one or another of the impurity atoms, but the interaction introduced by the assembly and the interaction of this assembly with the rest of the lattice and all this should be treated as a whole. As we deal with a crystal we are dealing with a many body problem [19]. It is well known that most of the salient features of local modes might be understood with simple models [20]. The theory of the effect of localized defects such as impurities, holes and interstitials on the vibrations of crystal lattices, which involve the use of Greens's functions, was developed independently by Lifshitz [21] and Montroll [22]. This theory [22] with a mass impurity model was applied by Lucovsky et. al. [23] and they were able to establish a large list for local mode frequencies which in many cases are quite close to observe modes and in other cases could serve as predictions. The agreement between reported and calculated frequencies was fair, differences been of the order of  6\% . More important, they observed that local mode frequencies as calculated by their technique and full lattice dynamics calculations differ at most 1\%  percent. The case of a pair silicon atoms in germanium, using the Montroll's theory [22] was done by Feldman et. al. [24] with a impurity model, which has two masses substitutionally introduced in an otherwise perfect crystal. Results of the calculation are in accord with experimental Raman studies.

The effects to be expected for two different impurities in the Montroll's lattice, have not been studied theoretically and this is the case which we present here. It is the purpose of this work to present results of one calculation for isotope frequency shifts of the BH local pair modes in crystalline silicon made as a many body problem with a simple model that are in accord with the experimental measurements. This work removes the troubling inconsistency above mentioned about the interaction between B$^{10}$ (B$^{11}$) and H(D) in silicon. In addition, it shows that both of two local modes and their isotope frequency shifts can clearly be understood.
\section{Previous calculations}
In silicon boron is substitutional [9,10,11] and it is presumed that hydrogen is interstitial [15,16]. Both atoms, boron and hydrogen, are lighter than silicon atoms. Frequencies for local single modes that are produce independently with their respective isotope shifts can be calculated using the simple model of masses which has nearest neighbour interactions [20,22,23]. These calculations of frequencies and isotope shifts have shown excellent results. With substitutional boron calculated frequencies are in accord with those obtained with models much more elaborated within 1\% [23]. The agreement between experimental values and calculated frequencies is also fair, differences are of the order of 6.5\%. In addition, it should be noted that for boron isotope frequency shift, the estimated value with this simple model is 8.33\% less than experimental measurement. However, this isotope frequency shift calculated with much more elaborated models is 4.2\% less or 8.33\% larger than the experimental value respectively [25,26]. Hence, differences obtained between each one of these models and experimental measurements are of the same order of magnitude for an estimation of isotope frequency shift. On the other hand, for interstitial hydrogen it is easy to show, using this simple model of masses [27], that differences between experimental values and calculated frequencies are of the order of 3.5\%. And for isotope frequency shift, it is only 1.8\% larger than the experimental value. Considering the above results, it seems that BH local pair mode frequencies, in crystalline silicon, might be calculated using this simple model of masses in Montroll's lattice with nearest neighbour interactions.
\section{Method}
Boron hydrogen pair configuration, in crystalline silicon, is yet controversial. Several sites have been proposed for the BH complex [1,2,3,4,5,6]. However, in this work it is assumed that boron is substitutional and hydrogen is interstitial in a BC site adjacent to boron. Local pair mode frequencies introduced by two identical substitutional adjacent masses in a linear lattice with nearest neighbour interaction have already been calculated using the Green's function method [22,28]. Also, other mathematical techniques were used for that [29]. In this paper, local pair mode frequencies due to two different adjacent masses, one substitutional and the other interstitial, in a lineal lattice of masses with nearest neighbour interactions, have been analytically calculated by using finite difference techniques [30]. The equation obtained by this exact method is
\begin{align}
&\epsilon_1 \epsilon_2 (1-\epsilon_1-\epsilon_2+\epsilon_1 \epsilon_2)x^{3} + 2 [\epsilon_1 \epsilon_2 (2-\epsilon_1 - \epsilon_2) +\epsilon_1 (\epsilon_1-1)+ \epsilon_2(\epsilon_2 - 1)]x^{2}\nonumber\\&+ [\epsilon_1(\epsilon_1-4) + \epsilon_2(\epsilon_2 - 4)+ 2\epsilon_1 \epsilon_2 + 3]x+4=0
\end{align}
where $\epsilon_1= m'/m$, $\epsilon_2= m''/m$, and $x= m \omega^{2}/K$. In Eq. (1), $m'$ is interstitial impurity mass, $m''$ is substitutional impurity mass, $m$ denotes the mass of atoms of the perfect lattice, K is the force constant of nearest neighbour interactions and $\omega$ indicates the angular frequency of the normal modes of vibration. Two local pair mode frequencies appear for both masses $m'$ and $m''$. In this many body problem the imperfection masses are the only interaction. It is easy to see that if $m'=m''$ in equation (1) then this gives the local pair mode frequencies obtained before [22,28,29].

When $m$ masses are chosen to be the masses of the real crystal then this model is of one parameter [20,23]. This parameter $K$ as usual is chosen to make the transverse optical phonon frequency of the host crystal to occur at 520 cm$^{-1}$. With these values local pair mode frequencies calculated using equation (1) are 1971.8 cm$^{-1}$ and 595.4 cm$^{-1}$ for the pair B$^{10}$H; 1969.3 cm$^{-1}$ and 577.6 cm$^{-1}$ for B$^{11}$H. Also, 1421.3 cm$^{-1}$ and 591.3 cm$^{-1}$ for B$^{10}$D and 1417.4 cm$^{-1}$ and 574.0 cm$^{-1}$ for B$^{11}$D.
\section{Discussion}
For boron local pair mode frequencies the results are 12\% lower than the experimental values. Hydrogen local pair mode frequencies are 3.5\% higher than experimental measurements. Note that differences between calculated and meassured frequencies, for BH local pair modes, are of the same order of magnitude than those found, with the mass model for boron and hydrogen local single mode frequencies ,in crystalline silicon. Also, it is easy to see that, for the large isotope frequency shifts these are almost 40\% shorter than the measured value for boron and 7.5\% larger than the experimental value for hydrogen. Hydrogen result is of the same order of magnitude than those in the case of a single impurity in silicon. However, for boron the difference is large. It appears that the influence of the lattice in this frequency range over boron local mode is strong. More important, calculations with this model show that splittings (small isotope frequency shifts) of hydrogen and deuterium lines with respect to the two boron isotopes are smaller than 3.9 cm$^{-1}$. These splittings are smaller than 3.3 cm$^{-1}$ in the experiment. The results for hydrogen and deuterium lines splittings are very accurate. In particular, hydrogen line splitting B$^{10}$H-B$^{11}$H was evaluated to be 2.5 cm$^{-1}$. This value is favourably compared with the experimental value of 0.8 cm$^{-1}$. For the line B$^{10}$D-B$^{11}$D the calculated value is 3.9 cm$^{-1}$. Comparison of this number with the experimental measurement 3.3 cm$^{-1}$ is excellent. Then, hydrogen and deuterium line splittings with B$^{10}$ and B$^{11}$ are very well reproduced by this model. There were no splitting in boron lines with H or D in the experiments, no for B$^{10}$ nor for B$^{11}$. Theoretically these splittings were calculated, with the model of this work, to be 4.6 cm$^{-1}$ for B$^{10}$H-B$^{10}$D and 3.6 cm$^{-1}$ for B$^{11}$H-B$^{11}$D. It should be remarked again that all the splittings calculated here are small (less or equal than 4.6 cm$^{-1}$). Nevertheless, it seems that is necessary to make a better resolution measurement at low temperature in order to see if these splittings for boron lines actually exist. Mainly because it is well known that there are numerous effects which shift the lines in the spectra by these small quantities [2,20,31,32]. From the theoretical point of view there is no reason why these small isotope frequency shifts, with mass change, should not exist. In general, obtained results for BH splittings are not only qualitative but quantitative in accord with the experiment. Finally, note how the above results indicate clearly that there are important interactions not only between the impurity masses B and H but also between them and the silicon lattice.

The first cluster theoretical calculation for H local pair mode of BH complex in crystalline silicon, gave a frequency at  1880 cm$^{-1}$ [33]. Another cluster calculation [34] gave two hydrogen frequencies at 1815 cm$^{-1}$ and at 810 cm$^{-1}$ for BH local pair modes. Distinct theoretical calculations performed at the density functional level, have reported 1820 cm$^{-1}$ in a bond center BC hydrogen position and 1000 cm$^{-1}$ in the antibonding AB hydrogen position for the H local pair mode frequency [35]. Also, frequencies at 1830 $\pm100$ cm$^{-1}$ BC, 1680 $\pm$ 100 cm$^{-1}$ AB and 1590 $\pm$ 100cm$^{-1}$ (back bonding position,BB), at the same level, for the H local pair mode have been obtained [36]. 1830cm$^{-1}$ was calculated in addition in BC configuration [37]. A more recent calculation [38] has given 2031 cm$^{-1}$ for the stretching mode frequency and 233 cm$^{-1}$ for the waggin mode frequency of the hydrogen. In this last case the H atom is located at the BC site of a Si-B bond. It should be noted that, in all of these theoretical calculations, there are large discrepancies between calculated frequencies and experimental values for the hydrogen local pair mode of the BH complex in crystalline silicon [39]. There are not numbers for deuterium and moreover, splittings of the hydrogen and deuterium lines have not been calculated. On the other hand, unfortunately there are not results for boron local pair mode frequency with these quantum chemical approaches which can be compared with experiments.

Using a phenomenological model [40] frequencies and isotope shift for hydrogen local pair mode of the BH complex in silicon, gave frequencies at 1870 cm$^{-1}$ for hydrogen and at 1360 cm$^{-1}$  for deuterium [41]. In that work it was assumed that  hydrogen atom position is the AB position. Although frequency values are an exact reproduction of experimental results for hydrogen, however,  this theoretical work did not predict the other boron local pair mode
and the splittings of the BH complex. Note that, this molecular model includes two parameters for the description of the perfect lattice. It has given good results for oxygen in GaP [40], nevertheless, now it uses
four parameters for the description of the BH impurity pair, so that the results for local pair mode frequencies obtained with it, are dependent of the choice of the values of the parameters. This was shown by another calculation in which frequencies are obtained for the BH local pair modes with the same model [42]. With a different adjustment of parameters, new frequencies were obtained at 1887 cm$^{-1}$  and 667 cm$^{-1}$ for the pair B$^{10}$H, at 1878 cm$^{-1}$ and 646 cm$^{-1}$ for B$^{11}$H, at 1409 cm$^{-1}$ and 529 cm$^{-1}$  for the pair B$^{10}$D and at 1395 cm$^{-1}$ and 528 cm$^{-1}$ for B$^{11}$D. However, now it is generally accepted that these results for isotope shifts and splittings are not in accord with experimental values. For example, it was mentioned before [17] that the above theoretical calculation [41] is inadequate, because it gives values for the splittings of the hydrogen and deuterium lines of 9 cm$^{-1}$  and 14 cm$^{-1}$, respectively. The shift B$^{10}$H-B$^{10}$D gives 138 cm$^{-1}$ and the shift B$^{11}$H-B$^{11}$D gives 118 cm$^{-1}$. In both cases the calculated values are very far of the experimental result [5,7,8]. Then for all these reasons it is clear that, this theoretical calculation [41] is not appropriated to explain the isotope shifts and the splittings of the BH complex in crystalline silicon. Notice, in addition, that all of the above microscopic models used to calculate frequencies of the BH pair modes, are of molecular character. Watkins et. al. [32] interpret the “anomalous” isotope shift in terms of the Fermi resonance between the second harmonic of the transverse boron vibration and the fundamental longitudinal deuterium vibration. Implicit in their treatment was the assumption that the B local mode has (E) symmetry. They also used perturbation theory in a molecular system. We believe this is a many body problem.
\section{Conclusion}
It was demonstrated that with this model of masses in a lattice which has nearest neighbour interactions, the two frequencies measured so far for BH local pair modes in crystalline silicon, can be reproduced. Furthermore, the large and small isotope frequency shifts calculated have shown an excellent agreement, not only qualitative but also quantitative with the experiment. It should be stressed, as well, that these results are not fortuitous. On the contrary, they are a direct consequence that for local pair mode frequency calculations the response of the entire lattice was taken into account. This, of course, was no sufficiently considered in models with molecular characteristics. Moreover, our calculations of isotope frequency shifts did not follow the current idea found in the literature that the presence of a second impurity in a complex may lead to a relatively small perturbation of one known situation [20,31]. More than that, here it was assumed that the second impurity causes a large perturbation, similar to that caused by the first impurity in the crystal and that it is the dynamics of the crystalline lattice and the imperfection itself which determine the frequencies of the local pair modes and all theirs isotope shifts. More experimental measurements with better resolution and at the same temperature (low) of the lines of the BH pair should as well be made to compare with more appropriated theoretical models.
\\
\\

\end{document}